\documentclass[fleqn,usenatbib]{mnras}

\usepackage{newtxtext,newtxmath}
\usepackage[T1]{fontenc}
\usepackage{xcolor}
\usepackage[normalem]{ulem}
\DeclareRobustCommand{\VAN}[3]{#2}
\let\VANthebibliography\thebibliography
\def\thebibliography{\DeclareRobustCommand{\VAN}[3]{##3}\VANthebibliography}

\usepackage{graphicx}	% Including figure files
\usepackage{amsmath}	% Advanced maths commands
\usepackage{comment}

\title[Anomalous relaxation in warm ion plasmas]{Anomalous thermal relaxation in warm ion plasmas}

\author[D. Barba-González et al.]{
D. Barba-González\thanks {E-mail: david.barbag@usal.es}
C. Albertus\thanks{E-mail: albertus@usal.es}
and M.A. Pérez-García\thanks{E-mail: mperezga@usal.es}
\\
Department of Fundamental Physics and IUFFyM,\\ Universidad de Salamanca, Plaza de la Merced s/n E-37008, Salamanca (Spain)
}

\date{Accepted XXX. Received YYY; in original form ZZZ}

\pubyear{2024}

\begin{document}
\label{firstpage}
\pagerange{\pageref{firstpage}--\pageref{lastpage}}
\maketitle

% Abstract of the paper
\begin{abstract} 
We perform microscopic simulations of the thermal relaxation of  warm neutral plasmas of astrophysical importance. Using Molecular Dynamics we study the thermal relaxation of a hot neutral fluid of finite-size neutron-rich ions  kept in a fixed-wall cool box. We show how the interplay among particle size, Yukawa interaction range and density are key to understand the features of the time-dependent thermal relaxation curve $T(t)$. We show that, under certain conditions, these systems exhibit faster cooling from increasingly larger initial temperature values. They also display non-Newtonian thermal behavior, including oscillations, that can be {\it effectively} interpreted as the consequence of the existence of a non-trivial system memory function. Finally, we consider the impact of multicomponent admixtures in the simulated system. We discuss these results and their possible extensions to astrophysical scenarios, where screened plasmas with Yukawa-like electrostatic potentials are usually involved.
\end{abstract}

% Select between one and six entries from the list of approved keywords.
% Don't make up new ones.

%%%%%%%%%%%%%%%%%%%%%%%%%%%%%%%%%%%%%%%%%%%%%%%%%%

%%%%%%%%%%%%%%%%% BODY OF PAPER %%%%%%%%%%%%%%%%%%

\section{Introduction}

\label{intro}
The behavior of non-equilibrium, relaxing systems is a subject of study that spans centuries and presumably was first recorded by Greek philosopher Aristotle circa 350 BC.  At much later times, around the late 17th century, Newton conducted the first experiments on the nature of cooling, noting that when the difference in temperature between two bodies is small, the rate of temperature equilibration was proportional to the temperature difference $ {\Delta T(t)}$. The dynamical equation  $
{d\Delta T(t)}/{dt} \propto - {\Delta T(t)}$ is solved yielding an exponential relaxation, usually sufficient to explain most idealized systems. However at larger temperature differences or more complex systems, it must be corrected. There is a wide variety of  experimental phenomena suggesting that a more general theory underlies these effects and as a complex system evolves out of equilibrium, its past history determines its fate \citep{Coleman1964}. 
\\
Equilibrium in thermodynamic systems is characterized by a loss of memory of previous history, and conversely, systems with broken ergodicity of some form are capable of retaining memory of their past history. However, this latter feature is not restricted to such systems, being also present in out-of-equilibrium conditions \citep{zwanPhysRev.124.983,2023PhRvL.130t7103T}. Generically, these systems behave in this particular way due to the {\it memory effects} and are known, alternatively, as {\it aging systems}, as the response to external perturbations from equilibrium configurations evolves in time depending on their previous history. One example of anomalous relaxation  is the Mpemba effect \citep{Mpemba1969}. It refers to the observation that the hotter of two identical beakers of water, put in contact exchanging heat with the same thermal reservoir, can cool faster under certain conditions. Note that although initially reported for water it is not specific to it and one can find this effect appearing in  granular matter \citep{pradosPhysRevLett.112.198001} and spin glasses \citep{BaityJesi2019}. Examples of amorphous solids from a binary mixture of particles under Lennard-Jones interactions also display memory effects \citep{fioccoPhysRevLett.112.025702}.
 At this point it is interesting to note that hysteresis effects, present in magnetized systems, can be understood as just another manifestation of Langevin dynamics in some contexts. In \cite{MAHATO1992220} a simple finite-volume Landau model based on magnetization $m(t)$ is studied to find supporting evidence for a theoretical hysteresis criterion based on rate competition. These ideas could have relevance in magnetic hysteresis more concretely in laser bistability and glass transition problems.

Apart from (low-density) terrestrial systems, thermal relaxation is important in the ultradense astrophysical context. Particular examples are electric interactions between dust and hot plasmas in the solar system \citep{LAFON199527}, the warm crust of  young proto NS being formed following the supernova explosion or heated after accreting material from a companion \citep{ArusPhysRevC.94.025805}. Ions are fully ionized while free electrons are the most mobile carriers of charge and typically ultradense warm matter cools from a few $\sim 10^{10}$ K (about temperature of 1 MeV, when we set $k_B=1$) to very low-temperature states, mostly due to neutrino emission, relaxing to an inhomogeneous ordered (or amorphous) configuration. 

It has been argued that the rapid cooling of the crust, during $\sim$ms hastily drops $T$, inhibiting a perfect lattice from forming in the inner NS crust. Previous works, see  \citep{pethickPhysRevLett.72.1964} or table 2 in the recent paper \citep{hut10.1093/mnras/stad967} have indicated that much of that cooling is due to neutrino emission in electron-nucleus bremsstrahlung or plasmon decay with an associated emissivity up to  $\sim 10^{33} T_8^6 M_{\mathrm{cr}} / M_{\odot} \mathrm{erg} \,\mathrm{s}^{-1}$, where $M_{\mathrm{cr}}$ is the mass of the crust and $T_8=T/10^8 K$. It is also important to remark that other sources such as phonon production and impurities are also important in this process. 
In this sense, calculations \citep{particles7040059} of thermal conductivity and thermal Hall effect in one-component electron-ion plasmas ($^{56}$Fe and $^{12}$C nuclei) at densities of outer crust of neutron stars (NS) and the interiors of white dwarfs up to thermal energies of $\sim$ 10 MeV, show that electron transport dominates, with electrons scattering off correlated nuclei via screened electromagnetic forces. In these  calculations plasma correlations in the liquid state are accounted for using an ion structure-function approach.

A complex composition due to particular behavior of ion heat capacity as obtained in a dynamical microscopic description \citep{barbamnras_10.1093/mnras/stae235} may occur in the bottom layers with high impurity parameter $Q_{\rm imp}$ displaying a mixture of different nuclei and thus an amorphous state \citep{PBJONESPhysRevLett.83.3589,potekhinrefId0}. 
Relying on rough estimates of exponential cooling, the estimated relaxation time is $\tau_{\rm NS \,crust}\sim 10^{-3}$ s. 

Superfluidity also plays a role in the disordered expected alloys \citep{sauls2020superfluidity}. The presence of pasta phases in the inner NS crust \citep{2018PhRvL.121m2701C} leads to inhomogeneity of cooling and quenching rates, far from the idealized conditions usually quoted. These systems are modeled from potentials involving competing (long- vs short-range) interactions due to their electromagnetic and hadronic nature. Yukawa-like potentials describe the in-medium screened electric spread charges of neutron-rich ions or massive mediators for nucleons \citep{barba22}.

In this context, Molecular Dynamics (MD) simulations have been shown to be a useful tool in the study of the microscopic behavior of astrophysical systems, especially in the study of their static and dynamical properties, such as diffusion \citep{2024PhRvL.133m5301C}, resistance to deformations in pasta phases \citep{2018PhRvL.121m2701C} or structure factor in the crystallized phase \citep{2016PhPl...23i2120D}. On a more sophisticated side, soft matter packages such as ESPResSo allow an efficient treatment of selected interactions (Lennard-Jones, Morse, Buckingham, Coulomb and others) for various geometries and 
 thermostats \citep{LIMBACH2006704}.

In this work, we perform MD simulations for systems that we first set to constant temperature evolution. After equilibration, we leave them to evolve out of equilibrium to study their thermal relaxation. From this, we find that for certain selected cases an ordered phase formation develops in the system. We do not intend to perform an exhaustive study of the phase space of the 3D plasma, but to display  some characteristic features showing how the screened finite-size ion plasma relaxes as it cools down. For this we solve the dynamics of a $N$ ion system inside a cubic box with volume $V=L^3$, allowing it transiting from a high-temperature to a low-temperature state through a collisional procedure of extraction of heat through contact with simulation box walls. As shown below, we consider electrostatic interactions and their consequence on relaxation, without including in this work the effect of magnetic field. As we explain, the obtained behavior in the fluid phase during thermal relaxation can be {\it effectively} understood as the result of the existence of a non-trivial memory function.

The paper is organized in the following manner. In Section 2 we present the general trends in cooling laws expected in  complex ion systems departing from the pure Newtonian case. We focus first on early cooling, where  stretched exponential temperature drop may exist to, later on, discuss oscillatory behavior due to the presence of complex-valued poles in the screened system velocity correlation function. We focus on finite-size and neutron-rich ions characterized by a generalized Coulomb theory parameter typical of warm crusts. This is of interest in warm materials in low-density phases of proto NS where electrons form a degenerate Fermi sea. Under these conditions we discuss the formation of crystallized states as obtained in our set of simulated cases that, although limited, are representative of the phase transition arising in Yukawa-like systems. We elaborate on the validity of our approximations.
In Section 3, we explain the implementation of the cooling procedure in the heat reservoir defined for our computational simulation box and obtain, in Section 4, the time-dependent $T(t)$ profiles with several initial temperatures, Yukawa spatial ranges and particle charge spread. We extract the velocity autocorrelation function $R(t)$ where the two differentiated regions of temperature drop show up, including the effect of the crystallization.  We discuss how these anomalous cooling patterns signal the presence of a non-trivial memory function. We further characterize the system by displaying the disruption of crystallized configurations when a mixture of ions with a lower $Z$ electric charge is considered.  We finally present our conclusions in Section 5.

\section{Thermal relaxation of an ion fluid}
\label{section2}

The simplest theoretical description for a weakly-interacting (classical) system that is relaxing thermally from one state into another is Newton's relaxation law, given by 

\begin{equation}
    \frac{d\Delta T(t)}{dt}=- \frac{\Delta T(t)}{\tau},
    \label{eq:newton}
\end{equation}

where $\Delta T(t)=T(t)-T_{\rm reservoir}$ is the difference between the  temperature of the system, $T(t)$, and the temperature of the reservoir (heat bath) with whom the system is in contact (exchanging heat).  $\tau$ is the relaxation time and the solution to Eq.(\ref{eq:newton}), assuming a time-independent $\tau$, is an exponential $\Delta T(t) = \Delta T (0)e^{-\frac{t}{\tau}}$.

However, in complex systems as those mentioned in the Introduction, $\tau$ may retain time dependence and induce anomalous non-monotonic $T(t)$, exhibiting oscillations \citep{2015JChPh.142j4106L}. As a result, the generalized stretched exponential relaxation at early times

\begin{equation}
\Delta T(t)= \Delta T (0)e^{- \left(\frac{t}{\tau}\right)^{\beta}}
    \label{eq:stretched}
\end{equation}
describes a dependence $\tau^{-1}\propto t^{\beta-1}$, which has been observed in more complex  systems, such as spin glasses. $\beta$ here is an free parameter that has been shown to have a temperature dependence during relaxation \citep{Phillips_1996}. Thus, in general, the microscopic behavior governs the $T$ decay, and delving deeper is crucial to understanding relaxation in realistic systems. 

In the original Langevin dynamics, originally coined for colloids, random terms were included to simulate the collisional forces that drive the motion of the particles in the system  \citep{vainstein2006}. Most importantly, the generalized Langevin equation  may incorporate the presence of a generic external potential \citep{fariasPhysRevE.80.031143}.

In the 3D plasma system we will focus in this work, we consider $N$ fully-ionized ions with positions  $\left\{r_i^k\right\}$, with $i=1, \ldots, N$ and $k=1, 2, 3$ placed in a box at fixed low temperature, $T_{\rm low}$. The system is prepared at an initial (high) temperature $T_{\rm init}$ and let dynamically evolve according to an interaction potential (including statistical noise). Heat extraction proceeds through wall collisions, as we explain below. We follow its thermal relaxation as it  approaches a target lower temperature $T_{\rm low}$. Such a system can be effectively thought as if dynamically evolving in presence of a memory function under the form 

\begin{equation}
m \ddot{r}_i^k (t)+\int_{t_0}^t d t^{\prime} \Gamma\left(t-t^{\prime}\right) \dot{r}_i^k \left(t^{\prime}\right)=-\frac{\delta V\left(\left\{\vec{r}_j\right\}\right)}{\delta r_i^k (t)}+\xi_i^k (t),
\label{langevin}
\end{equation}

where $t_0$ is some initial time, $\Gamma\left(t-t^{\prime}\right)$ is a dissipation kernel and the fluctuating force noise terms $\xi_i^k (t)$ with a non-zero two-point correlation \citep{zwanzig2001nonequilibrium}. $V\left(\left\{\vec{r}_j\right\}\right)$ is the interaction potential between ions.

More in detail, dissipation and stochastic (noise) terms are expected to originate from scattering events, thus giving rise to finite interaction times, that reflect in the system's equation of motion as non-local (i.e. non-Markovian) \citep{2015JChPh.142j4106L}. Thus, the total force has been partitioned into a systematic part and a fluctuating part (or noise). Both  come from the interaction of the particle with its dense environment.

In order to clarify the context where these Langevin dynamics appear, let us recall there are two basic views \citep{zwanzig2001nonequilibrium} of the nature of the fluctuating force $\xi_i^k (t)$. In the more-commonly presented view, the fluctuating force is supposed to come from occasional impacts among particles in the surrounding medium. The force during an impact is supposed to vary with extreme rapidity over the time of any observation, in fact, in any infinitesimal time interval. Then the effects of the fluctuating force can be summarized by giving its first and second moments, as time averages over an infinitesimal time interval. 
The other view can be illustrated by the analogy of random number generators in computers or error distribution in differential equation integration algorithms. A good algorithm is random in the sense that it satisfies various statistical requirements of randomness for almost all choices of seed or as step integration is infinitesimal. 

As mentioned, the system may have a memory described by $\Gamma(t)$ so that the frictional force at time $t$ is no longer determined just by a linear dissipative term but by an integral 
\begin{equation}
    \int_{-\infty}^t d t^{\prime} \Gamma(t-t^{\prime}) \dot{r}_i^k(t^{\prime}),
\end{equation}

over earlier times of this memory function \citep{vainstein2006}.

In order to set a physical scenario, in this work we consider a screened ion fluid so that the interaction potential $ V\left(\left\{\vec{r}_j\right\}\right)$ is given under Yukawa form, characterized by a spatial length $\lambda$. This is well motivated by charge screening, where the ion Coulomb potential $V_C\sim Ze/r$ becomes $V_Y\sim Ze^{-r/\lambda}/r$ when surrounded by an electron fluid according to the Debye-Hückel theory.  
For an ideal, point-like (PL) ion with charge  $+Z_i$  located at position $\vec{r}_i$ it reads
\begin{equation}
V^{\rm PL}_{Y,i}(\vec{r})=\frac{Z_i}{\left|\vec{r}-\vec{r}_i\right|} e^{-\frac{\left|\vec{r}-\vec{r}_i\right|}{\lambda}}.
\label{yuk}
\end{equation}

Instead, when considering more realistic finite-size ions the total potential adds individual contributions $V(\vec{r})=\sum_i V^{\rm GS}_i(\vec{r})$  obtained by convolving Eq. \eqref{yuk} with a  Gaussian-shaped (GS) charge in the form $\rho_{i, a_i}(r)=Z_i\left(\frac{a_i}{\pi}\right)^{\frac{3}{2}} e^{-a_i \left|\vec{r}-\vec{r}_i\right|^2}$. The associated charge width $a_i$ may depend on the particular species considered (as we will see later when discussing composition impact).  In this way, we obtain the dynamical equations governing the relaxing screened ion system as \citep{barbamnras_10.1093/mnras/stae235}
\begin{multline}
    m\ddot{r}^k_i+\int_{t_0}^t d t^{\prime} \Gamma\left(t-t^{\prime}\right) \dot{r}_i^k\left(t^{\prime}\right)=\sum_{j=1}^N    
    2\left(\frac{a_i}{\pi}\right)^{\frac{1}{2}}\frac{Z_i e^{-a_i r_{ij}^2}}{r_{ij}^2}\left(\frac{r_{ij}^k}{r_{ij}}\right)\times\\
    \left\{\int_0^{\infty} r^{\prime}V_{Z_i,a_i}\left(r^{\prime}\right)e^{-a_j r^{\prime 2}}\right.
    \bigg.\left[\left(1+2a_j r_{ij}^2\right)\sinh\left(2a_j r_{ij}^2\right)\right.  \\\left.-2a_j r_{ij}r^{\prime} \cosh{\left(2a_j r_{ij}^2\right)} \right]dr^{\prime}\bigg\}+\xi_i^k (t),
    \label{gle}
\end{multline}

where $r_{ij}=\left|\vec{r}_i-\vec{r}_j\right|$ $i,j=1,...,N$. $V^{\rm GS}_{i} (\vec{r})\equiv V_{Z_i,a_i}(\vec{r})$ is the Yukawa potential induced by a Gaussian charge density, explicitely
\begin{multline}
    V_{Z_i,a_i}\left(\vec{r}\right)=\frac{Z_i}{2|\vec{r}-\vec{r_i}|} e^{\frac{1}{4 a_i \lambda^2}}\left[e^{-\frac{|\vec{r}-\vec{r_i}|}{\lambda}}\mathrm{erfc}\left(\frac{1}{2\sqrt{a_i}\lambda} - \right.\right.\\ 
    \left.\left. \sqrt{a_i}|\vec{r}-\vec{r_i}|\right)  -e^{\frac{|\vec{r}-\vec{r_i}|}{\lambda}}\mathrm{erfc}\left(\frac{1}{2\sqrt{a_i}\lambda}+\sqrt{a_i}|\vec{r}-\vec{r_i}|\right) \right],
\end{multline}
where $\rm erfc$ is the complementary error function. 

We emphasize that the choice of this potential is due to its arisal in astrophysical scenarios, as the interaction between ions in the warm plasma in a proto-NS is screened by the presence of a polarizable degenerate electron sea \citep{barba22}. MD fluctuations inherent to this method play the role of the random $\xi_i^k (t)$. The memory function satisfies a self-consistent equation for the velocity autocorrelation function $C(t)=\langle v(t) v(0)\rangle$ under the form

\begin{equation}
    \frac{d R(t)}{d t}=-\int_0^t \Gamma\left(t-t^{\prime}\right) R\left(t^{\prime}\right) d t^{\prime},
\end{equation}
where the normalized correlation function is $R(t) \equiv C(t) /$ $C(0)$. This equation thus guides our understanding of the dynamical behaviour of the system in the fluid phase. We will numerically obtain $R(t)$ for the ion-screened system when relaxing as it cools down, see below.

More concretely, for relaxing phenomena the Laplace transform of the correlation function, $\mathcal{L}\{R(t)\}=\tilde{R}(s)$ is a good quantitative way of studying long-term behavior. In the Newtonian cooling, see Eq. \eqref{eq:newton}, the Laplace transform will be given by $\tilde{R}(s)=\frac{1}{s+1/\tau}$, displaying a single pole, $s=-1/\tau$ in the negative part of the real axis. However, when the behavior is oscillatory, complex-valued poles under the generic form $s=a+ib$ with non-vanishing $b\neq0$ appear.

To scrutinize in a more quantitative way the presence of long-term oscillatory behavior, right after the early cooling,  let us suppose the time correlation function $R(t)$  a combination of exponential and oscillatory functions under the form

\begin{equation}
R(t) =  d\left( \mathrm{e}^{-a_1 t} \mathrm{cos}\left(bt\right)+\mathrm{e}^{-a_2 t}\mathrm{sin}\left(bt\right)\right)+c
\label{eq:r_t}
\end{equation}

where $a_1,a_2,b,c,d$ are real positive quantities. This expression has a Laplace transform that can be written as

\begin{align}
\mathcal{L}\{R(t)\}=\tilde{R}(s)=\frac{c}{s}+\frac{d(a_1+s)}{\left(s+a_1-i b\right)\left(s+a_1+i b\right)}+\nonumber
\\
 \frac{b d}{\left(s+a_2-i b\right)\left(s+a_2+i b\right)}.
\label{eq:r_s}
\end{align}

By comparing with \cite{2015JChPh.142j4106L}
\begin{equation}
\tilde{R}(s)=\frac{1}{s+\tilde{\Gamma}(s)},
\end{equation}

we rewrite Eq. \eqref{eq:r_s} showing its explicit poles $s=-a_1 \pm i b, s=-a_2 \pm i b$ with

\begin{equation}
\tilde{\Gamma}(s)=\frac{1-c-s\left(\frac{d\left(a_1+s\right)}{b^2+\left(a_1+s\right)^2}+\frac{b d}{b^2+\left(a_2+s\right)^2}\right)}{\tilde{R}(s)} .
\end{equation}

Now in the absence of oscillations, i.e. the limit where $b\rightarrow 0$ we have

\begin{equation}
\lim _{b \rightarrow 0} \tilde{\Gamma}(s)= \frac{s(a_1+s)(1-c)-s^2d}{c(a_1+s)+sd} 
\quad s \neq-a_1,-a_2 ,
\end{equation}
so if we further simplify and set the irrelevant offset $c=0$ we obtain
%%%%%%%%%%%%%%%%%%%%%%%%%%%
\begin{figure}
    \centering   \includegraphics[width=0.5\textwidth]{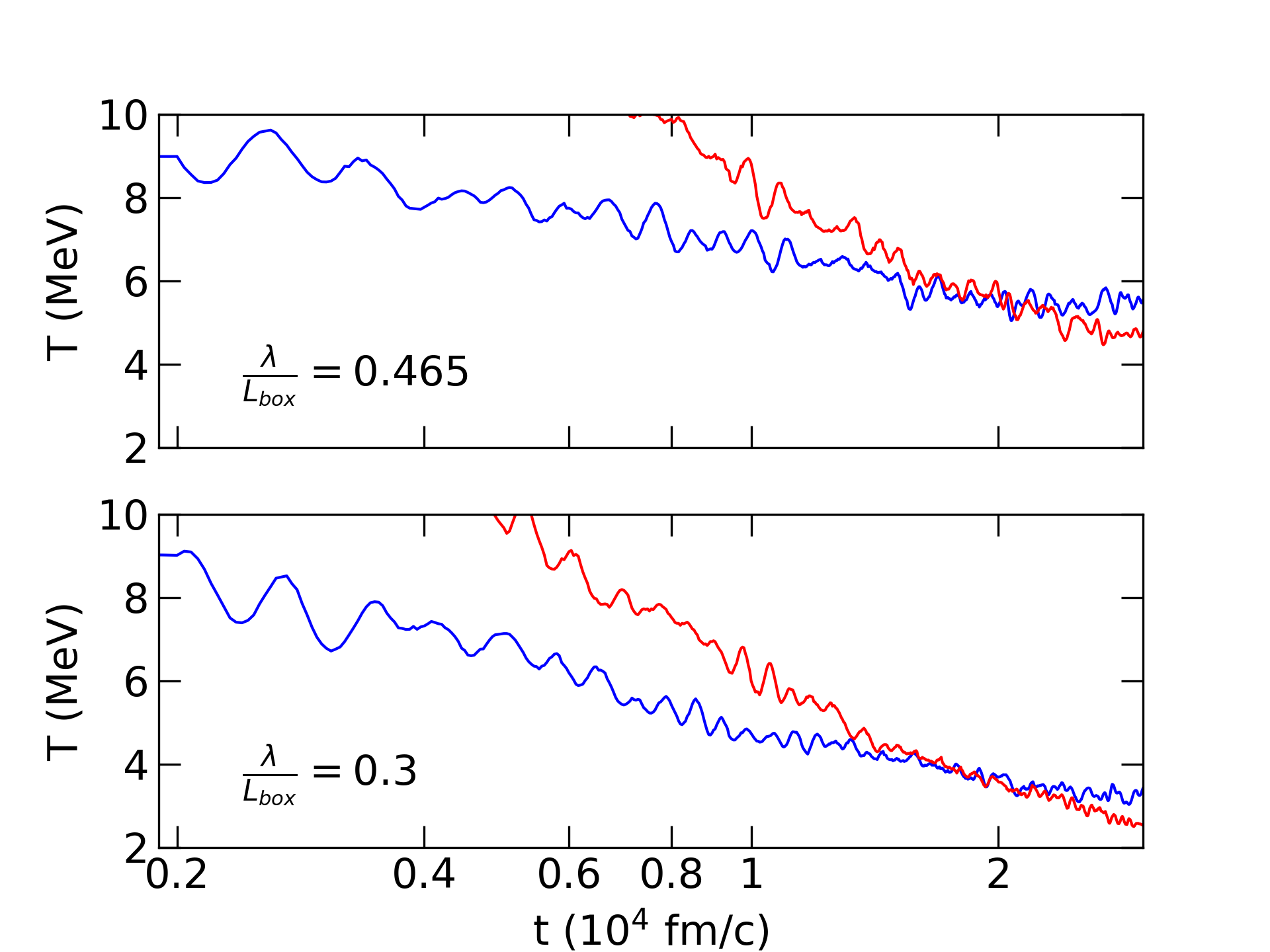}
    \caption{Temperature evolution from initially fixed temperature $T_{\rm init}=9$ MeV (blue) and $T_{\rm init}=15$ MeV (red) under Yukawa potential with screening length $\lambda/L=0.3$  (top) and  $\lambda/L = 0.465$  (bottom). We used ion species $(Z,A)=(38,128)$ at  $n_I = 1.1 \times 10^{-4} \rm \mspace{4mu}fm^{-3}$.} 
    \label{fig:lamL}
\end{figure}
%%%%%%%%%%%%%%%%%%%%%%%%%%
\begin{equation}
\tilde{\Gamma}(s)=\frac{\left(a_1+s(1-d)\right)}{d},
\end{equation}
so that the inverse transform ($d\neq0$) of $\tilde{R}(s)=d/(a_1+s)$ is 
\begin{equation}
\mathcal{L}^{-1}\{\tilde{R}(s)\}=d e^{-a_1 t},
\end{equation}

which corresponds to the memory function associated with Newtonian cooling $\sim e^{-a_1 t}$ with $\tau=\frac{1}{a_1}$ . 

In the same way, we can directly compare the early cooling from the usual Newtonian trend 

\begin{equation}
\mathcal{L}\left\{e^{- t/\tau}\right\}=\frac{1}{s+1/\tau}, \\
\end{equation}
to the stretched exponential 
\begin{equation}
\mathcal{L}\left\{e^{-(t / \tau)^\beta}\right\}=\frac{\tau}{s+\frac{1}{\tau^\beta} \Gamma\left(1+\frac{1}{\beta}\right)} ,
\end{equation}

where $\Gamma$ here is the mathematical function generalizing the factorial function. $\beta$ determines the drop, for $\beta<1$ it is slower than exponential while $\beta>1$ it is more rapid, instead. Clearly the density and temperature effects in the fluid interacting sample are driving the early cooling through $\beta$.

\section{Method}
\label{section3}
Simulations in this work are performed using our already well-tested MD codes \citep{barba22,barbamnras_10.1093/mnras/stae235} adapted to a cubic box of side length $L$. We thus simulate a Yukawa-interacting system of finite-spread charged particles set at initial (high) temperature $T_{\rm init}$ to later on start  tracking the microscopic ion dynamics using a suitable timestep much smaller than the ion plasma frequency  $\omega_{\mathrm{p}}=\left(4 \pi e^2 n_I Z^2 / m_{\mathrm{I}}\right)^{1 / 2}$ i.e. $dt\ll \omega^{-1}_p$ along its thermal relaxation. Here $m_{\rm I}$ and $n_{\rm I}$ are the ionic mass and number density, respectively, while $e$ is the electron charge and $Z$ the ion proton number.
In our simulations $N$ is kept constant as particles that get out of the box get refolded inside through the opposite side of the wall using periodic boundary conditions along with Minimum Image Convention in order to minimize the impact of finite-size effects. 

To induce the quenching effect, we implemented the  lowering of kinetic energy through the simulation box walls, which are assumed to have a temperature $T_{\rm wall}\equiv T_{\rm reservoir}$.  To cool the system down, we rescale the velocities of the $\delta N$ particles that get refolded by a factor $\sqrt{\frac{T_{\rm wall}}{T(t)}}$ where $T(t)$ is the ensemble-averaged instantaneous temperature of the sample. The efficiency of this procedure will be a factor less than unity that will arbitrarily rescale the relaxation times in a systematic way. We fix the wall temperature $T_{\rm wall}\sim 0.1$ MeV although we keep in mind that crystallization effects will prevent the fluid system from reaching such a temperature. 
Along these lines, a collisional, velocity-dependent friction term is analogously implemented in the study of granular and molecular mixture gases \citep{2022FrP....10.1671M, 2021PhFl...33e3301G,2023PhRvE.108b4902B}.

It is widely known that a screened Coulomb fluid  crystallizes as temperature $T$ drops beyond a critical value that for a classical one-component plasma in the weakly screened regime is $\Gamma_C\sim 171.8$ \citep{hamaguchiPhysRevE.56.4671} with tiny dependence on the fcc or bcc  lattices. This phase transition is characterized by the Coulomb plasma parameter  $\Gamma_C=\frac{Z^2}{lT}$ for  (point-like) ions with $Z$ charge and  mean distance $l$ at density $N/L^3$. Instead, as described in \cite{barba22} for a realistic bulk system, screening and finite size affect crystallization, so that it departs from the quoted  canonical value.

%%%%%%%%%%%%%%%%%%%%
\begin{figure}
    \centering
    \includegraphics[width=0.5\textwidth]{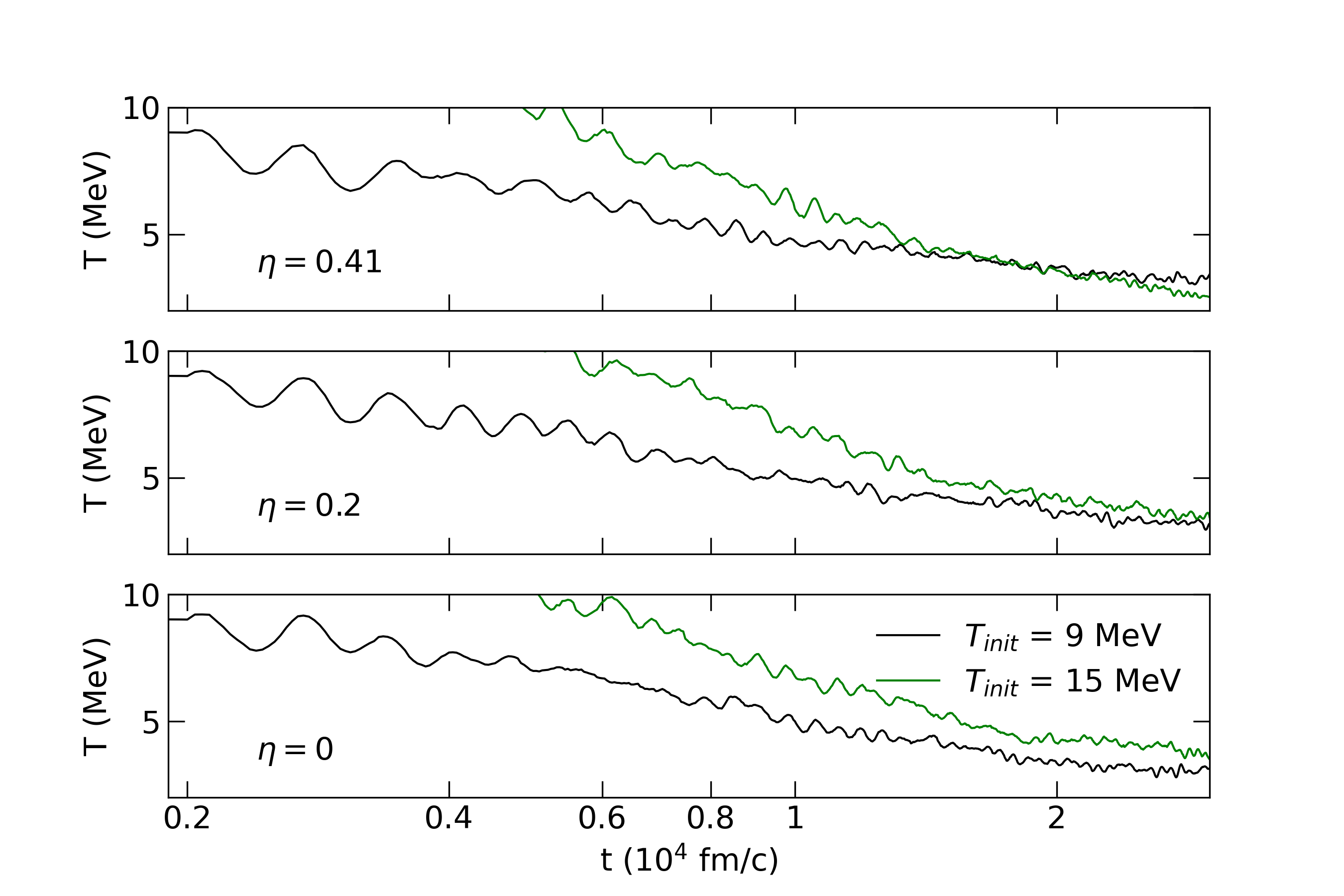}
    \caption{Temperature evolution for ions with finite charge spread  (top) $\eta =  0.41$, (middle) $\eta=0.2$ and bottom $\eta=0$ i.e. point-like particles. We set $T_{\rm init} = 9$ MeV (black) and  15 MeV (green) for ion species $(Z,A)=(38,128)$ at $n_I = 1.1 \times 10^{-4}$ $\rm fm^{-3}$ with $\lambda/L=0.3$.} 
    \label{fig:eta}
\end{figure}
%%%%%%%%%%%%%%%%%%%%%%%%%%

As already studied, the Debye screening length exhibits dependence in temperature and density \citep{hamaguchiPhysRevE.56.4671, potePhysRevE.65.036412}.  In this work we are interested in exploring thermally-relaxing Yukawa systems with interest in dense astrophysical plasmas where the screening length does not get largely distorted by thermal effects and approximate up to factors  of ${\sim \mathcal O}(1)$ to that associated to a degenerate electron fluid, i.e. the Thomas-Fermi screening length. Then  $\lambda \propto \lambda_e$, where $\lambda_e$ is the Thomas-Fermi screening length
$\lambda_e \sim \frac{1}{2 k_{F, e}} \sqrt{\frac{\pi}{\alpha}}$. $k_{F, e}$ is the electron Fermi momentum and $\alpha$ is the fine-structure constant. This is of interest for ions in astrophysical ultradense conditions such as those in warm materials in low-density phases of proto-NS where electrons form a degenerate Fermi sea. 
%%%%%%%%%%%%%%%%%%%%
\begin{figure*}
    \centering
    \includegraphics[width=1.0\textwidth]{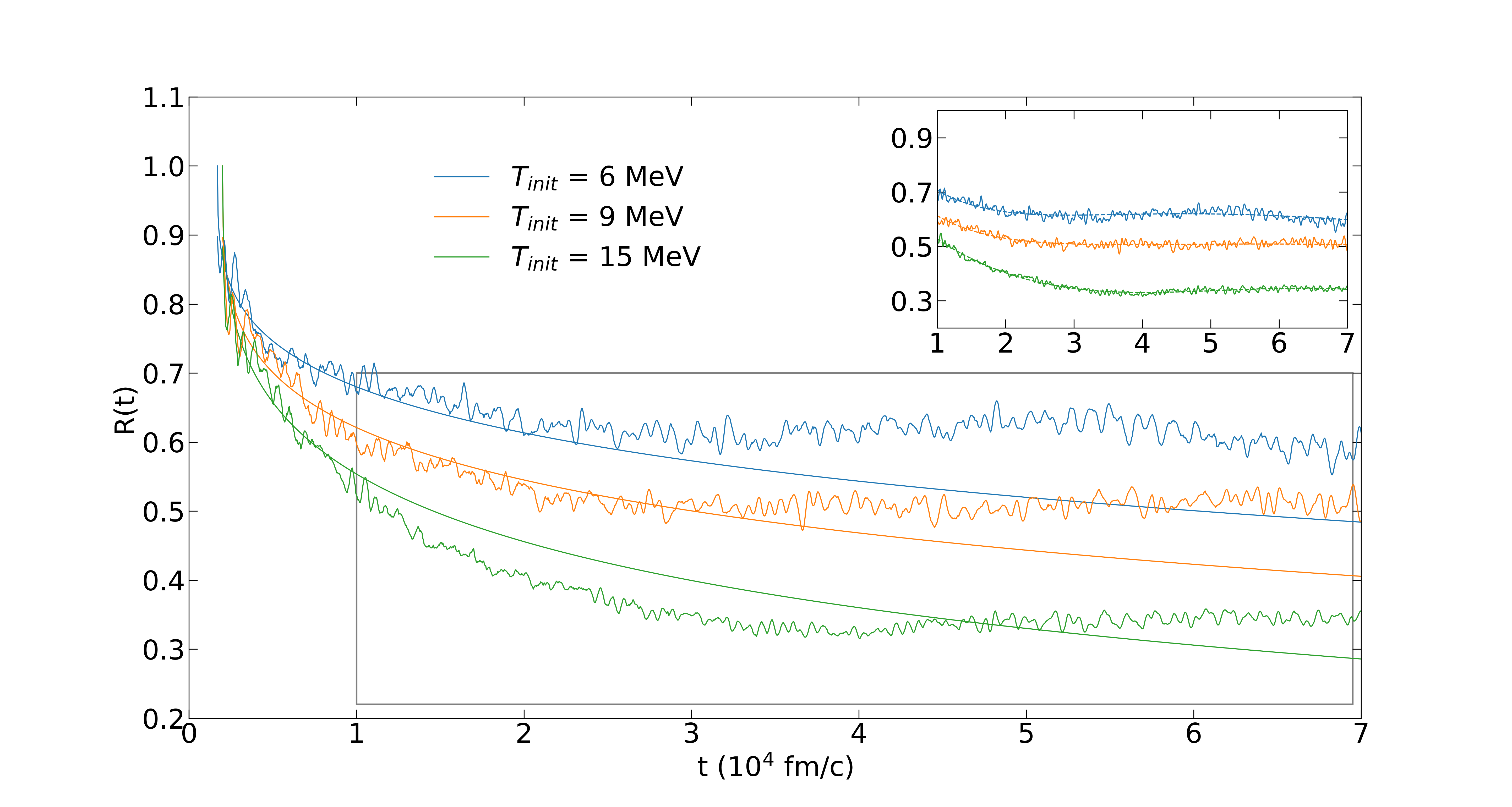}
    \caption{Normalized velocity autocorrelation function $R(t)$ for samples starting relaxation from $T_{\rm init} = 6, 9, 15$ MeV. Composition is fixed at $(Z,A)=(38,128)$, $\lambda/L=0.3$ and $\eta = 0.41$. In the large panel, smooth curves are fits to  stretched exponential form in Eq. ( \ref{eq:stretched}) during early relaxation. The inset shows the details of subsequent oscillating behavior due to ${\mathbb{C}}$-plane  poles in $\tilde{R}(s)$, see text for details.}
    \label{fig:stretched}
\end{figure*}
%%%%%%%%%%%%%%%%%%%%

In the simulation box we follow thermal relaxation including phase transitions (gas-liquid-solid) in the Yukawa fluid where the transition point in the density-$T$ phase space is highly dependent on the interaction screening length $\lambda$.

This is relevant to our results because the cooling of the sample will be affected by the liquid-solid phase transition, i.e. lattice formation. When the crystal appears the particles will stop crossing the simulation box's walls and thus the extraction of energy from the system comes to a halt
so that the sample will not reach $T_{\rm wall}$.
This means that at a particular timestep of the simulation, the average number of particles that are colliding with the walls is bounded by

\begin{equation}
    \delta N(t)<6 v(t) \Delta t L^2 n_I
    \label{rate}
\end{equation}
with $\Delta t$ the stepsize. In this contribution, we are simulating systems at ionic density $n_I = 1.1\times 10^{-4}$ $\rm fm^{-3}$, $N \lesssim 10^3$ where $\lambda_e\sim 21$ fm, following their thermal relaxation from initial temperatures $T_{\rm init}=$ 6, 9, 15 MeV. We will assume the warm fluid does not change in composition along the relaxation process, although we will comment on the effect of mixtures later on. 

We analyze, for the cases studied, how the thermal relaxation in the ion system proceeds, evaluating the monotony of the $T(t)$ curve and any possible anomalous crossing of the cooling curves depending on previous fluid history, starting from different $T_{\rm init}$. Apart from density and temperature, there are two relevant parameters that we have found to be crucial in determining characteristic features i.e. the charge spread of the particles, characterized \citep{barba22} by $\eta \equiv \eta_i= 1/\sqrt{a_i}l$ and the ratio $\lambda/L$. We remark at this point that we do not intend an exhaustive study on the phase-space of the fluid, but to evaluate specific trends present in these systems, leaving extensive analysis for a future contribution. 

In order to meaningfully compare our simulation results \citep{PhysRevLett.119.148001}, we start the thermal relaxation procedure from thermalized samples that reproduce a Maxwellian velocity distribution at $T_{\rm init}\equiv T(t_0)$. Later on, at time $t>t_0$, the Yukawa  system is allowed to relax in contact with the heat reservoir as prescribed, so that the $T(t)$ dependence on the rest of the parameters characterizing the system arises.

\section{Results and discussion}

Our MD code solves the dynamics for each individual particle in the setting above described imposing the cooling procedure and rescaling velocities for the $\delta N$ particles crossing box walls. We obtain, as an output, the positions and velocities of the ions as functions of time during the simulation allowing the reconstruction of the normalized autocorrelation function $R(t)$ during relaxation.  

In Fig. (\ref{fig:lamL}) we show the temperature of the sample $T(t)$ for initial temperatures 15 and 9 MeV, and two different screening lengths $\lambda=$70, 45.15 fm. They correspond to $\frac{\lambda}{ L}=$ 0.465 and 0.3, respectively at density $n_I = 1.1\times 10^{-4}\rm \mspace{4mu} fm^{-3}$. We set samples  $N\lesssim 10^3$, verifying our findings are not largely affected by finite size effects. Single composition is fixed with neutron-rich ion species $(Z,A)=(38,128)$ that we take as not far from typical of warm ultradense proto-NS crust environments \citep{pearson10.1093/mnras/sty2413}. We keep in mind that for temperatures larger than a few MeV fragmentation of nuclei must be taken into account. For the sake of simplicity we consider same species over the relaxation process. 

In these cooling curves, the fluid cools until the onset of crystallization, occurring at $t\sim 2\times 10^4$ fm/c. These screening lengths correspond to $\kappa=0.19, 0.29$, respectively, with $\kappa=\frac{\lambda}{l}$.
From top to bottom panels, $T$ crossing occurs, so that the initially hotter system cools down more rapidly than the initially cooler sample, as the phase transition arises before, halting the $T$ drop.  Thus we see that, in a sense, warmer samples can undergo more frequent scattering events with walls and efficiently achieve ordered configurations.

In Fig. (\ref{fig:eta}) we show the thermal evolution   dependence on the $\eta$ parameter and its effect on anomalous relaxation. As measured from electron-ion scattering and discussed in previous works \citep{barba22,barbamnras_10.1093/mnras/stae235} a Gaussian charge spread is assigned depending on the particular species. We take \cite{XurefId0} $a_i=\frac{3}{2\left\langle R^2\right\rangle}$ with a quadratic radius depending on the ion mass $A$, $\sqrt{\left\langle R^{2}\right\rangle}=\left(0.8 A^{1/3}+2.3\right)$. As shown, in  Fig. (\ref{fig:eta}) top panel depicts $\eta =  0.41$, middle $\eta=0.2$ and bottom $\eta=0$ i.e. point-like particles. We use the one-component ion system with species $(Z,A)=(38,128)$ at $n_I = 1.1 \times 10^{-4} \rm fm^{-3}$ representing a typical neutron-rich ion present in proto-NS matter approaching crystallized phases such as those in the crust \citep{pearson10.1093/mnras/sty2413}. It can be seen that the hot-cold crossing appears as $\eta$ grows, thus hinting at the crucial nature of the Gaussian spread of the charged particles and how neglecting this refinement when using point-like approximations ($\eta=0$) in the simulations may prevent from observing it.
%%%%%%%%%%%%%%%%%%%%%%%%%%%
\begin{figure*}
    \centering
    \includegraphics[width=1.0\textwidth]{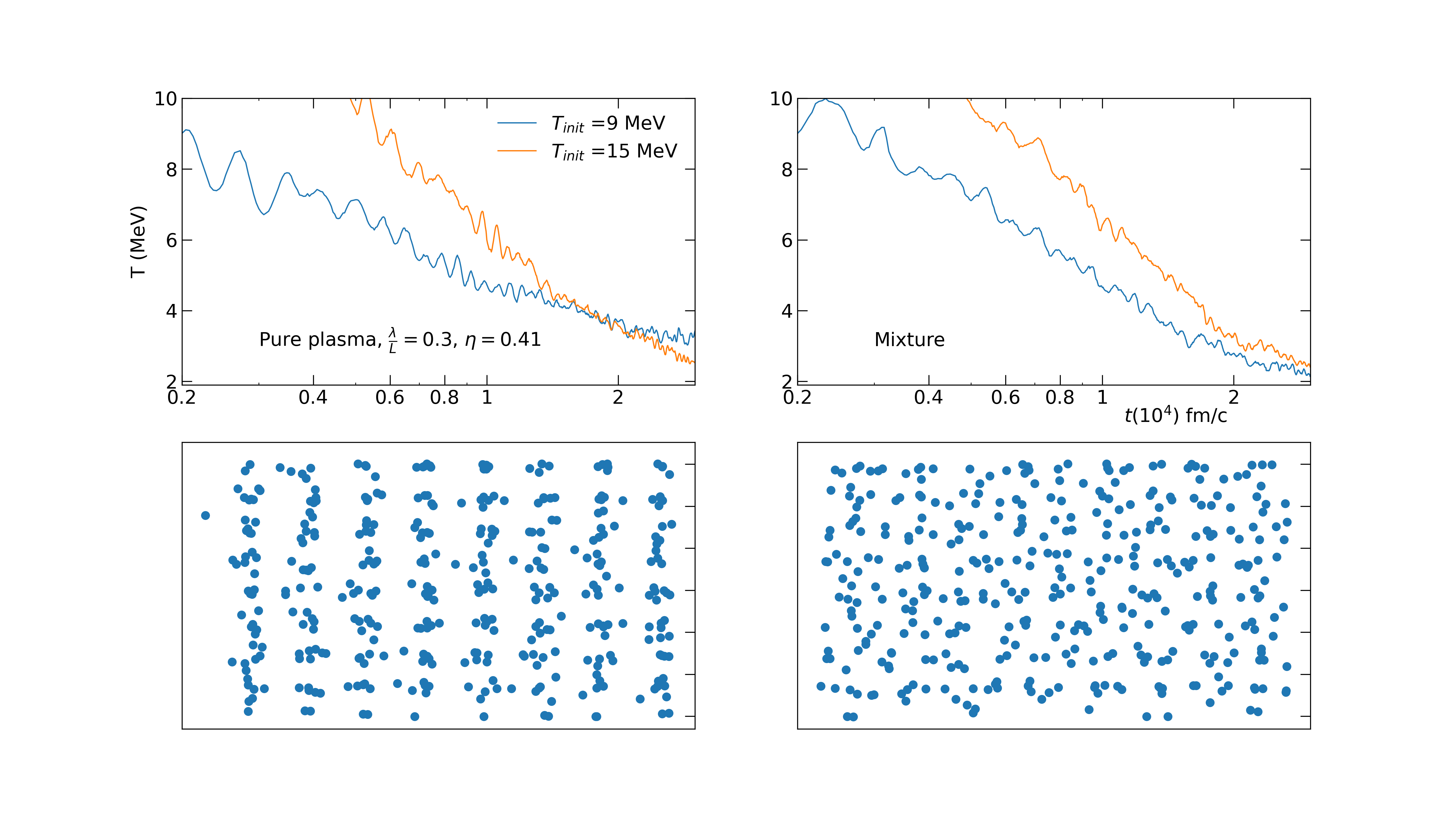}
    \caption{\textit{Top panel:} Temperature $T(t)$ for a pure sample with $(Z,A)=(38,128)$ (left) versus a mixture in which 40 \% of the original ions have been substituted by species $(Z,A)=(19,128)$. \textit{Bottom panel:} Snapshots of particle positions for the $T=9$ MeV runs at the same $t=2\times 10^4\rm fm/c$ where initially warmer (colder) curves cross,  showing in the left panel the ordered sample for the one-component versus the disordered state for the mixture.}%$\d
    \label{fig:mixture}
\end{figure*}
%%%%%%%%%%%%%%%%%%%%%%%%
To characterize the $T(t)$ behavior it is useful to analyze the time correlation function of the ion velocity, defined as     $C(t) = \langle v(t)v(0)\rangle$ where the brackets  $\langle ..\rangle$ represent the ensemble average. In particular, the normalized velocity correlation function, $R(t)=\frac{C(t)}{C(0)}$ informs of the diffusive process in the system. Note that these quantities have a quadratic dependence on the velocity of the particles, and thus they can be fitted to functional forms such as those in Eqs. (\ref{eq:newton}) and (\ref{eq:stretched}). $R(t)$ stores information about the memory of the initial state that the system keeps during relaxation. In fact, its time evolution can be extracted by ensemble averaging Eq. (\ref{langevin}), see \cite{2015JChPh.142j4106L}, thus showing that a departure from exponential behavior implies the presence of memory effects in the system. 

In particular, the time derivative of the correlation function $\beta=-2 \frac{d}{d t} \ln [R(t)]$ informs about the kind of relaxation followed. 
Stretched exponential relaxation then refers to cases where the correlation function decays as $\ln R(t) \propto-t^\beta$ with $\beta<1$. 
Newton cooling law corresponds to cases in which the correlation is $R(t)=\exp (-\gamma t)$, with $\beta=2 \gamma$, however, more generally $\beta(t)$ is a function of time.

As mentioned, the three-dimensional Yukawa fluid we are simulating undergoes a phase transition to a more ordered system during relaxation. Thus, in fluid early phases particles interact with the box at the expected rate $\delta N/\Delta t$, given by Eq. (\ref{rate}). In the long term due to competition among potential energy/thermal effects ordered lattice starts formation and the particles dramatically decrease scattering with the simulation box. 

In Fig. (\ref{fig:stretched}) we plot the normalized velocity autocorrelation function $R(t)$ for three runs starting from $T_{\rm init}=$ 6, 9, and 15 MeV. In the large panel se have fitted the early temperature drop  data  to a stretched exponential (solid smooth line) as stemming from Eq. (\ref{eq:stretched}). For our selected set of cases we find $\beta=0.38\pm 0.02$ while the relaxation times show some spread according to a distribution we label as $\rho(\tau)$ i.e. we find $\tau_6=118600 \pm 9300$ fm/c, $\tau_9 =  53600 \pm 3500$ fm/c, $\tau_{15} =  27750 \pm 990$ fm/c.  As shown in \cite{bouchaud2007anomalousrelaxationcomplexsystems}  any non trivial distribution of relaxation times $\rho(\tau)$  leads to a relaxation function that is faster than exponential on short times and slower than exponential on large times as we find for our system.

The depicted inset corresponds to the time window when the ordered phase is forming. From snapshots of positions  at $t=2\times 10^4\rm fm/c$  a lattice ordered state appears, see  Fig. \ref{fig:mixture} where the left top (bottom) panels display this feature. 

This departure from the stretched exponential behavior is usually quantified through the appearance of imaginary poles in the Laplace transform $\tilde{R}(s)$ of the normalized autocorrelation function $\tilde{R}(t)$ as explained in Section \ref{section2}. This behavior is shown in the curves in Fig. (\ref{fig:stretched}), where oscillations appear.

As already explained in Section \ref{section2}, we can phenomenologically characterize these oscillations fitting the velocity autocorrelation function $R(t)$  extracted from our data  to Eq. \eqref{eq:r_t} after the departure from the stretched exponential early cooling. Its Laplace transform $R(s)$ displays complex-valued poles $s_1,s_2$ and their conjugates with $b\neq0$ indicating anomalous oscillatory time evolution of the autocorrelation function $R(t)$. For the sake of illustration we find for the $T_{\rm in}=6$ MeV, that the poles lie at $s_1=(9.7\pm 0.3)\times10^{-5} \pm i(1.93\pm0.03)\rm \,c/fm$ and $s_2=(1.4\pm 0.1)\times10^{-5} \pm i(1.93\pm0.03)\rm\, c/fm$. Due to our restricted set of simulations at this point we do not intend to extract a full analysis of the $a_1,a_2$ with initial temperature leaving that for a future work. However, by inspection we can see in our scenario $a_1,a_2$ and $b$ are  decreasing functions or $T_{\rm init}$. Thus the larger  $T_{\rm init}$ is the more the pole leads towards the real values. These data and their fits to the oscillating autocorrelation functions are shown in the inset of Fig. \ref{fig:stretched} in the late cooling region of interest close to the crystallization phase transition.

Finally, we show the effect of the introduction of impurities in the sample in Fig. (\ref{fig:mixture}). 
The top left panel shows the $T(t)$ curves for a pure (one component) sample, the same case $(Z,A)=(38,128)$ where the temperature crossing appears, as previously discussed. When the sample is doped with an impurity $(Z,A)=(19,128)$ that entails 40$\%$ of the ions, the cooling curve $T(t)$ is largely affected and the hot-cold crossing completely disappears (top right). We deliberately choose two ion populations maintaining the same charge spread $a$ so that the non-crossing effect is genuinely due to the combination $Z/A$. As a result, the crystal ordered phase formation is delayed in time for the mixture and arises at a lower temperature (not shown), as expected from the lowering of the average ion charge. To illustrate, we depict (bottom panels) the spatially two-dimensional projection of the simulation box with the ordered phase appearing at simulation time $t=2\times10^4$ fm/c in a pure sample (left) versus the mixture (right).

%%%%%%%%%%%%%%%%%%%%%%%%%%%%%%%%%%%%%%%%%%
\section{Conclusions}
\label{section4}

In this contribution, we have studied the anomalous cooling of a sample of warm Yukawa fluid kept in a finite box with realistic finite-size ions including single- and multicomponent composition. We have focused on a warm plasma of neutron-rich ions as it is of interest for early crust formation in proto-NS. We have included early and long-term cooling allowing the system to transition from a fluid to a crystallized phase. Using MD microscopic many-body simulations, we have solved the dynamics of the ion system as it is driven out of equilibrium from an initially set temperature. The simulation box, which acts as a heat reservoir, allows the cooling of the sample driving it to a selected target lower temperature than the initial one, from a well-determined wall-scattering procedure. We follow the thermal relaxation process through the positions and velocities of individual (screened) interacting ions.
Our results show that temperature curve $T(t)$ displays an early time stretched exponential relaxation i.e. non-exponential temperature crossing, highly dependent on parameters of the system. Namely, the screening length of the Yukawa potential, $\lambda$, density and the associated width of the ionic charge density distributions, $\eta$. Effectively screened and largely spread ions facilitate the $T(t)$ crossing, thus initially hotter samples are cooling more rapidly than initially colder ones. We find that later on, close to crystallization, an oscillatory behavior arises due to the existence of complex-valued poles in the memory function $\Gamma(s)$. The amplitude of these oscillations seems to be a decreasing function of initial temperature offset i.e. largest initial difference $\Delta T(0)$ among the sample and heat reservoir. As obtained, the normalized velocity correlation function deviates from that with an exponential quenching, exposing the presence of memory effects in the system and non-monotonic $T(t)$ behavior. We have also shown that introducing impurities in the mixture, with smaller ionic charge, makes the crossing feature disappear.  We expect this effect could be present when the warm proto NS crust forms thus being important for induced non-monotonic changes in thermodynamical quantities critically dependent on temperature. These novel features may affect ultradense systems where Yukawa interactions naturally appear, such as screened astrophysical plasmas forming crystals in white dwarf cores, such as the recently challenging measurement for HD 190412 C \citep{venner10.1093/mnras/stad1719} and NS crusts or hydrodynamical electron-ion systems in laser cavities. More improvements in our modellization should be made before quantitative extraction of trends is made.

\section*{Acknowledgments}

D. B. G., C.A. and M.A.P.G. acknowledge partial support from the Spanish Ministry of Science PID2022137887NB-100, RED2022-134411-T, Junta de Castilla y Le\'on SA101P24, SA091P24 and RES resources under AECT-2023-1-0026, AECT-2024-2-0009 projects. D.B.G. acknowledges support from a Ph.D. Fellowship funded by Consejería de Educación de la Junta de Castilla y León and European Social Fund.

%%%%%%%%%%%%%%%%%%%%%%%%%%%%%%%%%%%%%%%%%%%%%%%%%%
\section*{Data Availability}

 The computed data presented and discussed in this paper will be shared upon reasonable request.

%%%%%%%%%%%%%%%%%%%% REFERENCES %%%%%%%%%%%%%%%%%%

% The best way to enter references is to use BibTeX:

\bibliographystyle{mnras}
\bibliography{bibfile} % if your bibtex file is called example.bib

% Alternatively you could enter them by hand, like this:
% This method is tedious and prone to error if you have lots of references
%\begin{thebibliography}{99}
%\bibitem[\protect\citeauthoryear{Author}{2012}]{Author2012}
%Author A.~N., 2013, Journal of Improbable Astronomy, 1, 1
%\bibitem[\protect\citeauthoryear{Others}{2013}]{Others2013}
%Others S., 2012, Journal of Interesting Stuff, 17, 198
%\end{thebibliography}

%%%%%%%%%%%%%%%%%%%%%%%%%%%%%%%%%%%%%%%%%%%%%%%%%%

%%%%%%%%%%%%%%%%% APPENDICES %%%%%%%%%%%%%%%%%%%%%

\appendix

% Don't change these lines
\bsp	% typesetting comment
\label{lastpage}
\end{document}